\documentclass[12pt]{article}
\usepackage{amsfonts}
\usepackage{amsmath}
\usepackage{amsthm}
\usepackage{multirow}
\usepackage{graphicx}
\usepackage{epstopdf}
\usepackage{caption}
\usepackage{subcaption}
\usepackage[right=2.1cm,left=2.1cm, top=2.5cm, bottom=2.5cm]{geometry}

\allowdisplaybreaks

\title{Efficient Bayesian inference for stochastic volatility models \\[2pt]
 with ensemble MCMC methods}
\author{Alexander Y. Shestopaloff \\
Department of Statistical Sciences \\
University of Toronto \\
alexander@utstat.utoronto.ca \\
\and Radford M. Neal \\
Department of Statistical Sciences \\
\& Department of Computer Science \\
University of Toronto \\
radford@utstat.utoronto.ca}
\date{9 December 2014}
\begin{document}

\maketitle

\begin{abstract}
In this paper, we introduce efficient ensemble Markov Chain Monte Carlo (MCMC) sampling methods for Bayesian computations in the univariate stochastic volatility model. We compare the performance of our ensemble MCMC methods with an improved version of a recent sampler of Kastner and Fruwirth-Schnatter (2014). We show that ensemble samplers are more efficient than this state of the art sampler by a factor of about $3.1$, on a data set simulated from the stochastic volatility model. This performance gain is achieved without the ensemble MCMC sampler relying on the assumption that the latent process is linear and Gaussian, unlike the sampler of Kastner and Fruwirth-Schnatter.
\end{abstract}

\noindent
The stochastic volatility model is a widely-used example of a state space model with non-linear or non-Gaussian transition or observation distributions. It models
observed log-returns $y = (y_{1}, \ldots, y_{N})$ of a financial time series with time-varying volatility, as follows:
\begin{eqnarray}
Y_{i} | x_{i} &\sim& N(0, \exp(c + \sigma x_{i})), \quad i = 1, \ldots, N \\
X_{1} &\sim& N(0, 1/(1 - \phi^{2})) \\
X_{i} | x_{i-1} &\sim& N(\phi x_{i-1}, 1)
\end{eqnarray}
Here, the latent process $x_{i}$ determines the unobserved log-volatility of $y_{i}$. Because the relation of the observations to the latent state is not linear and Gaussian, this model cannot be directly handled by efficient methods based on the Kalman filter.

In a Bayesian approach to this problem, we estimate the unknown parameters $\theta = (c, \phi, \sigma)$ by sampling from their marginal posterior distribution $p(\theta | y)$. This distribution cannot be written down in closed form. We can, however, write down the joint posterior of $\theta$ and the log-volatilities $x = (x_{1}, \ldots, x_{N})$, $p(\theta, x | y)$ and draw samples of $(\theta, x)$ from it. Discarding the $x$ coordinates in each draw will give us a sample from the marginal posterior distribution of $\theta$.

To sample from the posterior distribution of the stochastic volatility model, we develop two new MCMC samplers within the framework of ensemble MCMC, introduced by Neal (2010). The key idea underlying ensemble MCMC is to simultaneously look at a collection of points (an ``ensemble'') in the space we are sampling from, with the $K$ ensemble elements chosen in such a way that the density of interest can be simultaneously evaluated at all of the ensemble elements in less time than it would take to evaluate the density at all $K$ points separately. 

Previously, Shestopaloff and Neal (2013) developed an ensemble MCMC sampler for non-linear, non-Gaussian state space models, with ensembles over latent state sequences, using the embedded HMM (Hidden Markov Model) technique of Neal (2003), Neal et al. (2004). This ensemble MCMC sampler was used for Bayesian inference in a population dynamics model and shown to be more efficient than methods which only look at a single sequence at a time. In this paper we consider ensemble MCMC samplers that look not only at ensembles over latent state sequences as in Shestopaloff and Neal (2013) but also over a subset of the parameters. We see how well both of these methods work for the widely-used stochastic volatility model. 

\section{Bayesian inference for the stochastic volatility model}

Bayesian inference for the stochastic volatility model has been extensively studied. In this paper, we focus on comparisons with the method of Kastner and Fruwirth-Schnatter (2014). This state-of-the-art method combines the method of Kim et al. (1998) with the ASIS (Ancillary Sufficiency Interweaving Strategy) technique of Yu and Meng (2011). Kastner and Fruwirth-Schnatter's method consists of two parts. The first is an update of the latent variables $x$ and the second is a joint update of $\theta$ and the latent variables $x$. We improve this method here by saving and re-using sufficient statistics to do multiple parameter updates at little additional computational cost. 

\subsection{A linear Gaussian approximation for sampling latent sequences}

The non-linear relationship between the latent and the observation process prohibits the direct use of Kalman filters for sampling the latent variables $x_{i}$. Kim et al. (1998) introduced an approximation to the stochastic volatility model that allows using Kalman filters to draw samples of $x_{i}$ which can be later reweighed to give samples from their exact posterior distribution. This approximation proceeds as follows. First, the observation process for the stochastic volatility model is written in the form
\begin{eqnarray}
\log(y_{i}^2) = c + \sigma x_{i} + \zeta_{i}
\end{eqnarray}
where $\zeta_{i}$ has a $\log(\chi^{2}_{1})$ distribution. 

Next, the distribution of $\zeta_{i}$ is approximated by a ten-component mixture of Gaussians with mixture weights $\pi_{k}$, means $m_{k}$ and variances $\tau^{2}_{k}, k = 1, \ldots, 10$. The values of these mixture weights, means and variances can be found in Omori (2007). At each step of the sampler, at each time $i$, a single component of the mixture is chosen to approximate the distribution of $\zeta_{i}$ by drawing a mixture component indicator $r_{i} \in \{1, \ldots, 10\}$ with probabilities proportional to 
\begin{eqnarray}
P(r_{i} = k | y_{i}, x_{i}, c, \sigma) \propto \pi_{k}\frac{1}{\tau_{k}}\exp((\log(y_{i}^{2}) - c - \sigma x_{i})^{2}/\tau^{2}_{k})
\end{eqnarray}

Conditional on $r_{i}$, the observation process is now linear and Gaussian, as is the latent process:
\begin{eqnarray}
\log(Y_{i}^{2}) | x_{i}, r_{i}, c, \sigma &\sim& N(m_{r_{i}} + c + \sigma x_{i}, \tau_{r_{i}}^{2}), \quad i = 1, \ldots, N \\
X_{1} | \phi &\sim& N(0, 1/(1 - \phi^{2})) \\
X_{i} | x_{i-1} &\sim& N(\phi x_{i-1}, 1)
\end{eqnarray}

Kalman filtering followed by a backward sampling pass can now be used to sample a latent sequence $x$. For a description of this sampling procedure, see Petris et al. (2009).

The mis-specification of the model due to the approximation can be corrected using importance weights
\begin{eqnarray}
w^{(l)} = \frac{\prod_{i=1}^{N}f(y_{i} | x_{i}^{(l)}, c^{(l)}, \sigma^{(l)})}{\prod_{i=1}^{N}(\sum_{k=1}^{10}\pi_{k}g(\log(y_{i}^{2}) | x_{i}^{(l)}, c^{(l)}, \sigma^{(l)}, m_{k}, \tau_{k}))}
\end{eqnarray} 
where $f$ is the $N(0, \exp(c + \sigma x_{i}))$ density, $g$ is the $N(m_{k} + c + \sigma x_{i}, \tau_{k})$ density and the index $l$ refers to a draw. Posterior expectations of functions of $\theta$ can then be computed as $\sum w^{(l)} f(\theta^{(l)})$, with $\theta^{(l)}$ the draws.

We note that, when doing any other updates affecting $x$ in combination with this approximate scheme, we need to continue to use the same mixture of Gaussians approximation to the observation process. If an update drawing from an approximate distribution is combined with an update drawing from an exact distribution, neither update will draw samples from their target distribution, since neither update has a chance to reach equilibrium before the other update disturbs things. We would then be unable to compute the correct importance weights to estimate posterior expectations of functions of $\theta$.

\subsection{ASIS updates}

ASIS methods (Yu and Meng (2011)) are based on the idea of interweaving two parametrizations. For the stochastic volatility model, these are the so-called non-centered (NC) and centered (C) parametrizations. The NC parametrization is the one in which the stochastic volatility model was originally presented above. The C parametrization for the stochastic volatility model is
\begin{eqnarray}
Y_{i} | \tilde{x}_{1} &\sim& N(0, \exp(\tilde{x}_{i})), \quad i = 1, \ldots, N \\
\tilde{X}_{1} &\sim& N(c, \sigma^{2}/(1 - \phi^{2})) \\
\tilde{X}_{i} | \tilde{x}_{i-1} &\sim& N(c + \phi (\tilde{x}_{i-1} - c), \sigma^{2})
\end{eqnarray}
The mixture of Gaussians approximation for C is the same as for NC.

Kastner and Fruwirth-Schnatter (2014) propose two new sampling schemes, GIS-C and GIS-NC, in which they interweave these two parametrizations, using either the NC or C parameterization as the baseline. The authors report a negiligible performance difference between using NC or C as the baseline. For the purposes of our comparisons, we use the method with NC as the baseline, GIS-NC, which proceeds as follows.
\begin{enumerate}
\item
Draw $x$ given $\theta, r, y$ using the linear Gaussian approximation update (NC)

\item
Draw $\theta$ given $x, r, y$ using a Metropolis update (NC)

\item
Move to C by setting $\tilde{x} = c + \sigma x$

\item
Draw $\theta$ given $\tilde{x}, r, y$ using a Metropolis update (C)

\item
Move back to NC by setting $x = \frac{\tilde{x} - c}{\sigma}$

\item
Redraw the mixture component indicators $r$ given $\theta, x, y$. 
\end{enumerate}

Theorem 4 of Yu and Meng (2011) establishes a link between ASIS and the PX-DA (Parameter Expansion-Data Augmentation) method of Liu and Wu (1999). In the case of the stochastic volatility model, this means that we can view the ASIS scheme for updating $x$ and $\theta$ as a combination of two updates, both done in the NC parametrization. The first of these draws new values for $\theta$ conditional on $x$. The second draws new values for both $x$ and $\theta$, such that when we propose to update $c$ to $c^{*}$ and $\sigma$ to $\sigma^{*}$, we also propose to update the sequence $x$ to $x^{*} = ((c + \sigma x) - c^{*})/\sigma^{*}$. For this second update, the Metropolis acceptance probability needs to be multiplied by a Jacobian factor $(\sigma/\sigma^{*})^{N}$ to account for scaling $\sigma$. A joint translation update for $c$ and $x$ has been previously considered by Liu and Sabatti (2000) and successfully applied to to stochastic volatility model. Scale updates are considered by Liu and Sabatti (2000) as well, though they do not apply them to the stochastic volatility model.

The view of ASIS updates as joint updates to $\theta$ and $x$ makes it easier to see why ASIS updates improve efficiency. At first glance, they look like they only update the parameters, but they actually end up proposing to change both $\theta$ and $x$ in a way that preserves dependence between them. This means that moves proposed in ASIS updates are more likely to end up in a region of high posterior density, and so be accepted.

Kastner and Fruwirth-Schnatter (2014) do a single Metropolis update of the parameters for every update of the latent sequence. However, we note that given values for the mixture indices $r$, $y$ and $x$, low-dimensional sufficient statistics exist for all parameters in the centered parametrization. In the non-centered parametrization, given $r$, $y$ and $x$, low-dimensional sufficient statistics exist for $\phi$. We propose doing multiple Metropolis updates given saved values of these sufficient statistics (for all parameters in the case of C and for $\phi$ in the case of NC). This allows us to reach equilibrium given a fixed latent sequence at little computational cost since additional updates have small cost, not dependent on $N$. Also, this eliminates the need to construct complex proposal schemes, since with these repeated samples the algorithm becomes less sensitive to the particular choice of proposal density.

The sufficient statistics in the case of NC are
\begin{eqnarray}
t_{1} = \sum_{i=1}^{N}x_{i}^{2}, \quad t_{2} = \sum_{i=2}^{N}x_{i-1}x_{i}, \quad t_{3} = x_{1}^{2} + x_{N}^{2}
\end{eqnarray}

with the log-likelihood of $\phi$ as a function of the sufficient statistics being
\begin{eqnarray}
\log(L(\phi | t)) = (1/2)\log(1 - \phi^{2}) - (1/2)(\phi^{2}(t_{1} - t_{3}) - 2\phi t_{2} + t_{1}) \nonumber
\end{eqnarray}

In the case of C the sufficient statistics are
\begin{eqnarray}
\tilde{t}_{1} = \sum_{i=1}^{N}\tilde{x}_{i}^{2}, \quad \tilde{t}_{2} = \sum_{i=2}^{N-1}\tilde{x}_{i}^{2} \quad \tilde{t}_{3} = \sum_{i=2}^{N}\tilde{x}_{i-1}\tilde{x}_{i} \quad \tilde{t}_{4} = \sum_{i=2}^{N-1}\tilde{x}_{i} \quad \tilde{t}_{5} = \tilde{x}_{1} + \tilde{x}_{N}
\end{eqnarray}

with the log-likelihood as a function of the sufficient statistics being
\begin{eqnarray}
\log(L(c, \phi, \sigma^{2} | \tilde{t})) &=& -(N/2)\log(\sigma^{2}) + (1/2)\log(1 - \phi^{2}) \\ \nonumber
&&-(1/2)(\tilde{t}_{1} + \phi^{2}\tilde{t}_{2} - 2\phi \tilde{t}_{3} - 2c\phi^{2} \tilde{t}_{4} - 2c(\tilde{t}_{4} + \tilde{t}_{5}) \\ \nonumber
&&+ 4c\phi \tilde{t}_{4} + 2c\phi\tilde{t}_{5} + (N-1)(c(\phi -1))^{2} + c^{2}(1-\phi^{2}))/\sigma^{2} \nonumber
\end{eqnarray}

The details of the derivations are given in the Appendix.

\section{Ensemble MCMC methods for \\ stochastic volatility models}

The general framework underlying ensemble MCMC methods was introduced by Neal (2010). An ensemble MCMC method using embedded HMMs for parameter inference in non-linear, non-Gaussian state space models was introduced by Shestopaloff and Neal (2013). We briefly review ensemble methods for non-linear, non-Gaussian state space models here.

Ensemble MCMC builds on the idea of MCMC using a temporary mapping. Suppose we are interested in sampling from a distribution with density $\pi(z)$ on $\mathcal{Z}$. We can do this by constructing a Markov chain with transition kernel $T(z'|z)$ with invariant distribution $\pi$. The temporary mapping strategy takes $T$ to be a composition of three stochastic mappings. The first mapping, $\hat{T}$, takes $z$ to an element $w$ of some other space $\mathcal{W}$. The second, $\bar{T}$, updates $w$ to $w'$. The last, $\check{T}$, takes $w'$ back to some $z' \in \mathcal{Z}$. The idea behind this strategy is that doing updates in an intermediate space $\mathcal{W}$ may allow us to make larger changes to $z$, as opposed to doing updates directly in $\mathcal{Z}$. 

In the ensemble method, the space $\mathcal{W}$ is taken to be the $K$-fold Cartesian product of $\mathcal{Z}$. First, $z$ mapped to an ensemble $w = (z^{(1)}, \ldots, z^{(K)})$, with the current value $z$ assigned to $z^{(k)}$, with $k \in \{1, \ldots, K\}$ chosen uniformly at random. The remaining elements $z^{(j)}$ for $j \neq k$ are chosen from their conditional distribution under an ensemble base measure $\zeta$, given that $z^{(k)} = z$. The marginal density of an ensemble element $z^{(k)}$ in the ensemble base measure $\zeta$ is denoted by $\zeta(z^{(k)})$. Next, $w$ is updated to $w'$ using any update that leaves invariant the \textit{ensemble density}
\begin{eqnarray}
\rho(w) = \rho((z^{(1)}, \ldots, z^{(K)})) = \zeta((z^{(1)}, \ldots, z^{(K)}))\frac{1}{K}\sum_{i=1}^{K}\frac{\pi(z^{(k)})}{\zeta_{k}(z^{(k)})}
\end{eqnarray}
Finally, a new value $z'$ is chosen by selecting an element $z^{(k)}$ from the ensemble with probabilities proportional to $\pi(z^{(k)})/\zeta_{k}(z^{(k)})$. The benefit of doing, say Metropolis, updates in the space of ensembles is that a proposed move is more likely to be accepted, since for the ensemble density to be large it is enough that the proposed ensemble contains at least some elements with high density under $\pi$. 

In Shestopaloff and Neal (2013), we consider an ensemble over latent state sequences $x$. Specifically, the current state, $(x, \theta)$, consisting of the latent states $x$ and the parameters $\theta$ is mapped to an ensemble $y = ((x^{(1)}, \theta), \ldots, (x^{(K)}, \theta))$ where the ensemble contains all distinct sequences $x^{(k)}$ passing through a collection of pool states chosen at each time $i$. The ensemble is then updated to $y' = ((x^{(1)}, \theta'), \ldots, (x^{(K)}, \theta')$ using a Metropolis update that leaves $\rho$ invariant. At this step, only $\theta$ is changed. We then map back to a new $x' = (x', \theta')$, where $x'$ is now potentially different from the original $x$. We show that this method considerably improves sampling efficiency in the Ricker model of population dynamics.

As in the original Neal (2010) paper, we emphasize here that applications of ensemble methods are worth investigating when the density at each of the $K$ elements of an ensemble can be computed in less time than it takes to do $K$ separate density evaluations. For the stochastic volatility model, this is possible for ensembles over latent state sequences, and over the parameters $c$ and $\sigma^{2}$. In this paper, we will only consider joint ensembles over $x$ and over $\sigma$. Since we will use $\eta = \log(\sigma^{2})$ in the MCMC state, we will refer to ensembles over $\eta$ below.

We propose two ensemble MCMC sampling schemes for the stochastic volatility model. The first, ENS1, updates the latent sequence, $x$, and $\eta$ by mapping to an ensemble composed of latent sequences $x$ and values of $\eta$, then immediately mapping back to new values of $x$ and $\eta$. The second, ENS2, maps to an ensemble of latent state sequences $x$ and values of $\eta$, like ENS1, then updates $\phi$ using an ensemble density summing over $x$ and $\eta$, and finally maps back to new values of $x$ and $\eta$.

For both ENS1 and ENS2, we first create a pool of $\eta$ values with $L_{\eta}$ elements, and at each time, $i$, a pool of values for the latent state $x_{i}$, with $L_{x}$ elements. The current value of $\eta$ is assigned to the pool element $\eta^{[1]}$ and for each time $i$, the current $x_{i}$ is assigned to the pool element $x_{i}^{[1]}$. (Since the pool states are drawn independently, we don't need to randomly assign an index to the current $\eta$ and the current $x_{i}$'s in their pools.) The remaining pool elements are drawn independently from some distribution having positive probability for all possible values of $x_{i}$ and $\eta$, say $\kappa_{i}$ for $x_{i}$ and $\lambda$ for $\eta$.  

The total number of ensemble elements that we can construct using the pools over $x_{i}$ and over $\eta$ is $L_{\eta}L_{x}^{N}$. Naively evaluating the ensemble density presents an enormous computational burden for $L_{x} > 1$, taking time on the order of $L_{\eta}L_{x}^{N}$. By using the forward algorithm, together with a ``caching'' technique, we can evaluate the ensemble density much more efficiently, in time on the order of $L_{\eta}L_{x}^{2}N$. The forward algorithm is used to efficiently evaluate the densities for the ensemble over the $x_{i}$. The caching technique is used to efficiently evaluate the densities for the ensemble over $\eta$, which gives us a substantial constant factor speed-up in terms of computation time.

In detail, we do the following. Let $p(x_{1})$ be the initial state distribution, $p(x_{i} | x_{i-1})$ the transition density for the latent process and $p(y_{i} | x_{i}, \eta)$ the observation probabilities. We begin by computing and storing the initial latent state probabilities --- which do not depend on $\eta$ --- for each pool state $x_{1}^{[k]}$ at time $1$.
\begin{eqnarray}
P_{1} = (p(x_{1}^{[1]}), \ldots, p(x_{1}^{[L_{x}]}))
\end{eqnarray}

For each $\eta^{[l]}$ in the pool and each pool state $x_{1}^{[k]}$ we then compute and store the initial forward probabilities
\begin{eqnarray}
\alpha_{1}^{[l]}(x_{1}^{[k]} | \eta^{[l]}) &=& p(x_{1}^{[k]})\frac{p(y_{1}|x_{1}^{[k]}, \eta^{[l]})}{\kappa_{1}(x_{1}^{[k]})}
\end{eqnarray}
Then, for $i > 1$, we similarly compute and store the matrix of transition probabilities
$$ P_{i} = 
\begin{pmatrix}
p(x_{i}^{[1]}|x_{i-1}^{[1]}) & \ldots & p(x_{i}^{[L_{x}]}|x_{i-1}^{[1]}) \\
\vdots & \ddots & \vdots \\
p(x_{i}^{[1]}|x_{i-1}^{[L_{x}]}) & \ldots & p(x_{i}^{[L_{x}]}|x_{i-1}^{[L_{x}]}) \\
\end{pmatrix}
$$
where
\begin{eqnarray}
p(x_{i}^{[k_{1}]} | x_{i-1}^{[k_{2}]}) \propto \exp(-(x_{i}^{[k_{1}]} - \phi x_{i-1}^{[k_{2}]})^{2}/2)
\end{eqnarray}
are transition probabilities between pool states $x_{i-1}^{[k_{2}]}$ and $x_{i}^{[k_{1}]}$ for $k_{1}, k_{2} \in \{1, \ldots, L_{x}\}$. We then use the stored values of the transition probabilities $P_{i}$ to efficiently compute the vector of forward probabilities for all values of $\eta^{[l]}$ in the pool
\begin{eqnarray}
\alpha_{i}^{[l]}(x_{i} | \eta^{[l]}) &=& \frac{p(y_{i}|x_{i}, \eta^{[l]})}{\kappa_{i}(x_{i})}\sum_{k=1}^{L_{x}} p(x_{i} | x_{i-1}^{[k]}) \alpha_{i-1}^{[l]}(x_{i-1}^{[k]} | \eta^{[l]}), \quad i = 1, \ldots, N
\end{eqnarray}
with $x_{i} \in \{x_{i}^{[1]}, \ldots, x_{i}^{[L_{x}]}\}$.

At each time $i$, we divide the forward probabilities $\alpha_{i}^{[l]}(x_{i})$ by $c_{i}^{[l]} = \sum_{k=1}^{L_{x}}\alpha_{i}^{[l]}(x_{i} | \eta^{[l]})$, storing the $c_{i}^{[l]}$ values and using the normalized $\alpha_{i}^{[l]}$'s in the next step of the recursion. This is needed to prevent underflow and for ensemble density computations. In the case of all the forward probabilities summing to $0$, we set the forward probabilities at all subsequent times to $0$. Note that we won't get underflows for all values of $\eta^{[l]}$, since we are guaranteed to have a log-likelihood that is not $-\infty$ for the current value of $\eta$ in the MCMC state. 

For each $\eta^{[l]}$, the ensemble density can then be computed as
\begin{eqnarray}
\rho^{[l]} = \prod_{i=1}^{N} c_{i}^{[l]}
\end{eqnarray}
To avoid overflow or underflow, we work with the logarithm of $\rho^{[l]}$.

Even with caching, computing the forward probabilities for each $\eta^{[l]}$ in the pool is still an order $L_{x}^{2}$ operation since we multiply the vector of forward probabilities from the previous step by the transition matrix. However, if we do not cache and re-use the transition probabilities $P_{i}$ when computing the forward probabilities for each value of $\eta^{[l]}$ in the pool, the computation of the ensemble densities $\rho^{[l]}$, for all $l = 1, \ldots, L_{\eta}$, would be about $10$ times slower. This is because computing forward probabilities for a value of $\eta$ given saved transition probabilities only involves multiplications and additions, and not exponentiations, which are comparatively more expensive.

In ENS1, after mapping to the ensemble, we immediately sample new values of $\eta$ and $x$ from the ensemble. We first sample a $\eta^{[l]}$ from the marginal ensemble distribution, with probabilities proportional to $\rho^{[l]}$. After we have sampled an $\eta^{[l]}$, we sample a latent sequence $x$ conditional on $\eta^{[l]}$, using a stochastic backwards recursion. The stochastic backwards recursion first samples a state $x_{N}$ from the pool at time $N$ with probabilities proportional to $\alpha_{N}^{[l]}(x_{N} | \eta^{[l]})$. Then, given the sampled value of $x_{i}$, we sample $x_{i-1}$ from the pool at time $i-1$ with probabilities proportional to $p(x_{i} | x_{i-1})\alpha_{i-1}^{[l]}(x_{i-1} | \eta^{[l]})$, going back to time $1$.

In the terminology of Shestopaloff and Neal (2013) this is a ``single sequence'' update combined with an ensemble update for $\eta$ (which is a ``fast'' variable in the terminology of Neal (2010) since recomputation of the likelihood function after changes to this variable is fast given the saved transition probabilities). 

In ENS2, before mapping back to a new $\eta$ and a new $x$ as in ENS1, we perform a Metropolis update for $\phi$ using the ensemble density summing over all $\eta^{[l]}$ and all latent sequences in the ensemble, $\sum_{l=1}^{L_{\eta}}\rho^{[l]}$. This approximates updating $\phi$ using the posterior density of $\theta$ with $x$ and $\eta$ integrated out, when the number of pool states is large.  The update nevertheless leaves the correct distribution exactly invariant, even if the number of pool states is not large.

\subsection{Choosing the pool distribution}

A good choice for the pool distribution is crucial for the efficient performance of the ensemble MCMC method. 

For a pool distribution for $x_{i}$, a good candidate is the stationary distribution of $x_{i}$ in the AR(1) latent process, which is $N(0, 1/\sqrt{1 - \phi^{2}})$. The question here is how to choose $\phi$. For ENS1, which does not change $\phi$, we can simply use the current value of $\phi$ from the MCMC state, call it $\phi_{\textnormal{cur}}$ and draw pool states from $N \sim (0, c/\sqrt{1 - \phi_{\textnormal{cur}}^2})$ for some scaling factor $c$. Typically, we would choose $c > 1$ in order to ensure that for different values of $\phi$, we produce pool states that cover the region where $x_{i}$ has high probability density.

We cannot use this pool selection scheme for ENS2 because the reverse transition after a change in $\phi$ would use different pool states, undermining the proof via reversibility that the ensemble transitions leave the posterior distribution invariant. However, we can choose pool states that depend on both the current and the proposed values of $\phi$, say $\phi$ and $\phi^{*}$, in a symmetric fashion. For example, we can propose a value $\phi^{*}$, and draw the pool states from $N \sim (0, c/\sqrt{1 - \phi_{\textnormal{avg}}^2})$ where $\phi_{\textnormal{avg}}$ is the average of $\phi$ and $\phi^{*}$. The validity of this scheme can be seen by considering $\phi^{*}$ to be an additional variable in the model; proposing to update $\phi$ to $\phi^{*}$ can then be viewed as proposing to swap $\phi$ and $\phi^{*}$ within the MCMC state. 

We choose pool states for $\eta$ by sampling them from the model prior. Alternative schemes are possible, but we do not consider them here. For example, it is possible to draw local pool states for $\eta$ which stay close to the current value of $\eta$ by running a Markov chain with some desired stationary distribution $J$ steps forwards and $L_{\eta} - J - 1$ steps backwards, starting at the current value of $\eta$. For details, see Neal (2003).

In our earlier work (Shestopaloff and Neal (2013)), one recommendation we made was to consider pool states that depend on the observed data $y_{i}$ at a given point, constructing a ``pseudo-posterior'' for $x_{i}$ using data observed at time $i$ or in a small neighbourhood around $i$. For the ensemble updates ENS1 and ENS2 presented here, we cannot use this approach, as we would then need to make the pool states also depend on the current values of $c$ and $\eta$, the latter of which is affected by the update. We could switch to the centered parametrization to avoid this problem, but that would prevent us from making $\eta$ a fast variable. 

\section{Comparisons}

The goal of our computational experiments is to determine how well the introduced variants of the ensemble method compare to our improved version of the Kastner and Fruwirth-Schnatter (2014) method. We are also interested in understanding when using a full ensemble update is helpful or not. 

\subsection{Data}

\begin{figure}[t]
         \centering
         \begin{subfigure}[b]{0.48\textwidth}
                 \includegraphics[width=\textwidth]{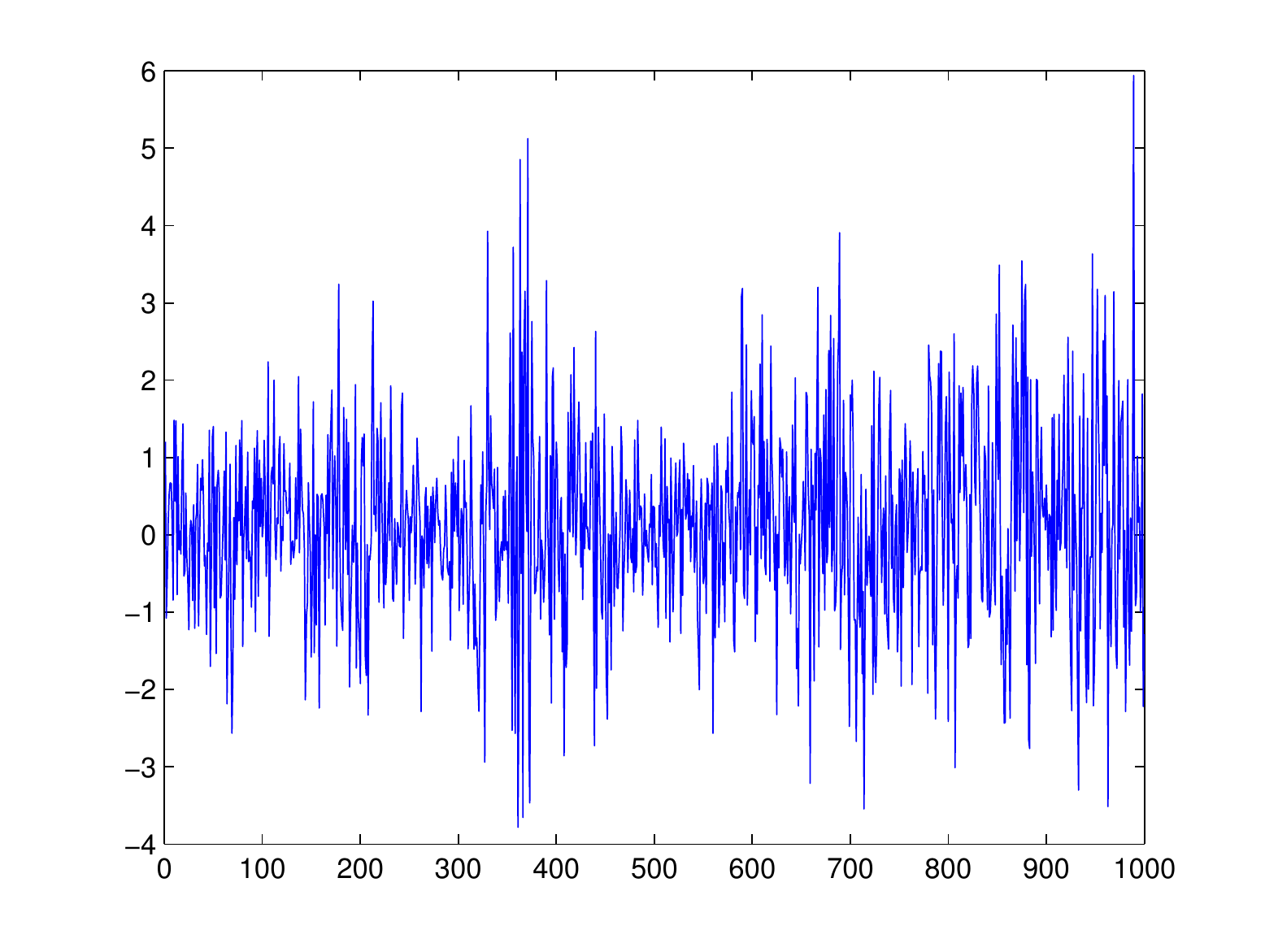}
                 \caption{y}
         \end{subfigure}
	 ~
         \begin{subfigure}[b]{0.48\textwidth}
                 \includegraphics[width=\textwidth]{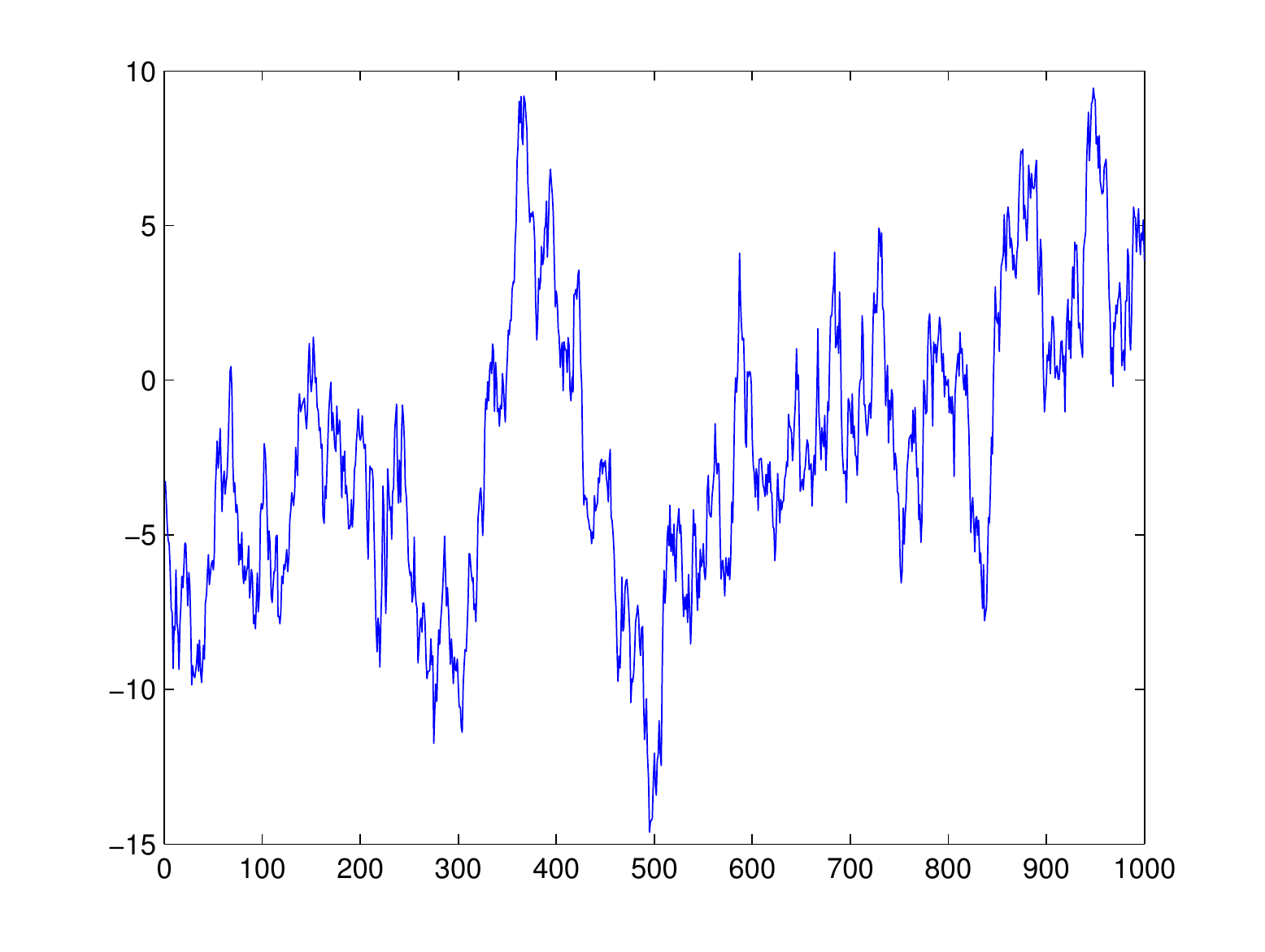}
                 \caption{x}
         \end{subfigure}
         \caption{Data set used for testing.}
         \label{fig:data}
\end{figure}

We use a series simulated from the stochastic volatility model with parameters $c = 0.5, \phi = 0.98, \sigma^{2} = 0.15$ with $N = 1000$. A plot of the data is presented in Figure \ref{fig:data}.

We use the following priors for the model parameters. 
\begin{eqnarray}
c &\sim& N(0, 1) \\
\phi &\sim& \textnormal{Unif}[0, 1] \\
\sigma^{2} &\sim& \textnormal{Inverse-Gamma}(2.5, 0.075)
\end{eqnarray}
We use the parametrization in which the Inverse-Gamma$(\alpha, \beta)$ has probability density
\begin{eqnarray}
f(x) = \frac{\beta^{\alpha}}{\Gamma(\alpha)}x^{\alpha-1}e^{-\beta/x}, \quad x > 0
\end{eqnarray}
For $\alpha = 2.5, \beta = 0.075$ the $2.5\%$ and $97.5\%$ quantiles 
of this distribution are approximately \\ $(0.0117, 0.180)$.

In the MCMC state, we transform $\phi$ and $\sigma^{2}$ to
\begin{eqnarray}
\eta &=& \log(\sigma^{2}) \\
\gamma &=& \log((1 + \phi)/(1 - \phi))
\end{eqnarray}
with the priors transformed correspondingly.

\subsection{Sampling schemes and tuning}

We compare three sampling schemes --- the Kastner and Fruwirth-Schnatter (KF) method, and our two ensemble schemes, ENS1, in which we map to an ensemble of $\eta$ and $x$ values and immediately map back, and ENS2, in which we additionally update $\gamma$ with an ensemble update before mapping back. 

We combine the ensemble scheme with the computationally cheap ASIS Metropolis updates. It is sensible to add cheap updates to a sampling scheme if they are available. Note that the ASIS (or translation and scale) updates we use in this paper are generally applicable to location-scale models and are not restricted by the linear and Gaussian assumption. 

Pilot runs showed that $80$ updates appears to be the point at which we start to get diminishing returns from using more Metropolis updates (given the sufficient statistics) in the KF scheme. This is the number of Metropolis updates we use with the ensemble schemes as well.

\noindent The KF scheme updates the state as follows:
\begin{enumerate}
\item 
Update $x$, using the Kalman filter-based update, using the current mixture indicators $r$.

\item 
Update the parameters using the mixture approximation to the observation density. This step consists of $80$ Metropolis updates to $\phi$ given the sufficient statistics for NC, followed by one joint update of $c$ and $\eta$.

\item 
Change to the C parametrization.

\item 
Update all three parameters simultaneously using $80$ Metropolis updates, given the sufficient statistics for C. Note that this update does not depend on the observation density and is therefore exact. 

\item 
Update the mixture indicators $r$.
\end{enumerate}

\noindent The ENS1 scheme proceeds as follows:
\begin{enumerate}
\item
Map to an ensemble of $\eta$ and $x$. 

\item
Map back to a new value of $\eta$ and $x$.

\item
Do steps 2) - 4) as for KF, but with the exact observation density. 
\end{enumerate}
\noindent The ENS2 scheme proceeds as follows:
\begin{enumerate}
\item
Map to an ensemble of $\eta$ and $x$. 

\item
Update $\gamma$ using an ensemble Metropolis update. 

\item
Map back to a new value of $\eta$ and $x$.

\item
Do steps 2) - 4) as for KF, but with the exact observation density. 
\end{enumerate}

The Metropolis updates use a normal proposal density centered at the current parameter values. Proposal standard deviations for the Metropolis updates in NC were set to estimated marginal posterior standard deviations, and to half of that in C. This is because in C, we update all three parameters at once, whereas in NC we update $c$ and $\eta$ jointly and $\phi$ separately. The marginal posterior standard deviations were estimated using a pilot run of the ENS2 method. The tuning settings for the Metropolis updates are presented in Table \ref{table:metstat}. 

For ensemble updates of $\gamma$, we also use a normal proposal density centered at the current value of $\gamma$, with a proposal standard deviation of $1$, which is double the estimated marginal posterior standard deviation of $\gamma$. The pool states over $x_{i}$ are selected from the stationary distribution of the AR(1) latent process, with standard deviation $2/\sqrt{1 - \phi_{\textnormal{cur}}^{2}}$ for the ENS1 scheme and $2/\sqrt{1 - \phi_{\textnormal{avg}}^{2}}$ for the ENS2 scheme. We used the prior density of $\eta$ to select pool states for $\eta$.

\begin{table}[b]
\small
\centering
\def\arraystretch{1.2}
\begin{tabular}{|c|*{3}{c}|c|c|*{3}{c}|c|}
\hline
\multirow{2}{*}{Method} & \multicolumn{3}{c|}{Prop. Std. (NC)} & Acc. Rate & Acc. Rate & \multicolumn{3}{c|}{Prop. Std. (C)} & Acc. Rate \\
			& $c$ & $\gamma$ & $\eta$ & for $\gamma$ (NC)& for $(c, \eta)$ (NC) & $c$ & $\gamma$ & $\eta$ & for $(c, \gamma, \eta)$ \\
\hline
KF   & \multirow{3}{*}{0.21} & \multirow{3}{*}{0.5} & \multirow{3}{*}{0.36} & \multirow{3}{*}{0.52} & 0.12 & \multirow{3}{*}{0.105} & \multirow{3}{*}{0.25} & \multirow{3}{*}{0.18} & \multirow{3}{*}{0.22} \\
\cline{6-6}
ENS1 & & & & & \multirow{2}{*}{0.12} & & & & \\
ENS2 & & & & & & & & & \\
\hline
\end{tabular}
\caption{Metropolis proposal standard deviations with associated acceptance rates.}
\label{table:metstat}
\end{table}

For each method, we started the samplers from $5$ randomly chosen points. Parameters were initalized to their prior means (which were $0$ for $c$, $1.39$ for $\gamma$ and $-3.29$ for $\eta$), and each $x_{i}, i = 1, \ldots, N$, was initialized independently to a value randomly drawn from the stationary distribution of the AR(1) latent process, given $\gamma$ set to the prior mean. For the KF updates, the mixture indicators $r$ where all initialized to $5$'s, this corresponds to the mixture component whose median matches the median of the $\log(\chi^{2}_{1})$ distribution most closely. All methods were run for approximately the same amount of computational time.

\subsection{Results}

Before comparing the performance of the methods, we verified that the methods give the same answer up to expected variation by looking at the $95\%$ confidence intervals each produced for the posterior means of the parameters. These confidence intervals were obtained from the standard error of the average posterior mean estimate over the five runs. The KF estimates were adjusted using the importance weights that compensate for the use of the approximate observation distribution. No significant disagreement between the answers from the different methods was apparent. We then evaluated the performance of each method using estimates of autocorrelation time, which measures how many MCMC draws are needed to obtain the equivalent of one independent draw.

To estimate autocorrelation time, we first estimated autocovariances for each of the five runs, discarding the first $10\%$ of the run as burn-in, and plugging in the overall mean of the five runs into the autocovariance estimates. (This allows us to detect if the different runs for each method are exploring different regions of the parameter/latent variable space). We then averaged the resulting autocovariance estimates and used this average to get autocorrelation estimates $\hat{\rho}_{k}$. Finally, autocorrelation time was estimated as $1 + 2\sum_{k=1}^{K}\hat{\rho}_{i}$, with $K$ chosen to be the point beyond which the $\rho_{k}$ become approximately $0$. All autocovariances were estimated using the Fast Fourier Transform for computational efficiency. 

The results are presented in Tables $\ref{table:ens1act}$ and $\ref{table:ens2act}$. The timings for each sampler represent an average over $100$ iteratons (each iteration consisting of the entire sequence of updates), with the samplers started from a point taken after the sampler converged to equilibrium. The program was written in MATLAB and run on a Linux system with an Intel Xeon X5680 3.33 GHz CPU. For a fair comparison, we multiply estimated autocorrelation times by the time it takes to do one iteration and compare these estimates.

\begin{table}[p]
\small
\centering
\def\arraystretch{1.2}
\begin{tabular}{|c|c|c|c|*{3}{c}|*{3}{c}|}
\hline
\multirow{2}{*}{$L_{x}$} & \multirow{2}{*}{$L_{\eta}$} & \multirow{2}{*}{Iterations} & \multirow{2}{*}{Time/iter (s)} & 
\multicolumn{3}{c|}{ACT} & \multicolumn{3}{c|}{ACT $\times$ time}  \\[-5pt]
 & & & 		& $c$ & $\gamma$ & $\eta$ & $c$ & $\gamma$ & $\eta$ \\
\hline
\multirow{4}{*}{10} &     1 &   195000 &     0.11 &      2.6 &       99 &      160 &     0.29 &       11 &       18 \\
	 	 &       10 &   180000 &     0.12 &      2.7 &       95 &      150 &     0.32 &       11 &       18 \\
	 	 &       30 &   155000 &     0.14 &      2.6 &       81 &      130 &     0.36 &       11 &       18 \\
	 	 &       50 &   140000 &     0.16 &      2.3 &       91 &      140 &     0.37 &       15 &       22 \\
\hline 
\multirow{4}{*}{30} &     1 &   155000 &     0.14 &      2.4 &       35 &       71 &     0.34 &      4.9 &      9.9 \\
	 	 &       10 &   135000 &     0.16 &      2.2 &       18 &       26 &     0.35 &      2.9 &      4.2 \\
	 	 &       30 &   110000 &     0.20 &      2.3 &       19 &       26 &     0.46 &      3.8 &      5.2 \\
	 	 &       50 &    65000 &     0.33 &      1.9 &       16 &       24 &     0.63 &      5.3 &      7.9 \\
\hline 
\multirow{4}{*}{50} &     1 &   115000 &     0.19 &      1.9 &       34 &       68 &     0.36 &      6.5 &       13 \\
	 	 &       10 &    95000 &     0.23 &      1.9 &       11 &       17 &     0.44 &      2.5 &      3.9 \\
	 	 &       30 &    55000 &     0.38 &      2.2 &      8.9 &       12 &     0.84 &      3.4 &      4.6 \\
	 	 &       50 &    55000 &     0.39 &      1.9 &       11 &       14 &     0.74 &      4.3 &      5.5 \\
\hline 
\multirow{4}{*}{70} &     1 &    85000 &     0.25 &      2.2 &       33 &       67 &     0.55 &      8.3 &       17 \\
	 	 &       10 &    60000 &     0.38 &      1.9 &      8.3 &       11 &     0.72 &      3.2 &      4.2 \\
	 	 &       30 &    50000 &     0.42 &      1.8 &      8.4 &       11 &     0.76 &      3.5 &      4.6 \\
	 	 &       50 &    45000 &     0.48 &      1.9 &      9.1 &       12 &     0.91 &      4.4 &      5.8 \\
\hline
\end{tabular}
\caption{Performance of method ENS1.}
\label{table:ens1act}
\end{table}

\begin{table}[p]
\vspace{4pt}
\small
\centering
\def\arraystretch{1.2}
\begin{tabular}{|c|c|c|c|c|*{3}{c}|*{3}{c}|}
\hline
\multirow{2}{*}{$L_{x}$} & \multirow{2}{*}{$L_{\eta}$} & \multirow{2}{*}{Acc. Rate for $\gamma$} & \multirow{2}{*}{Iterations} & \multirow{2}{*}{Time/iter (s)} & 
\multicolumn{3}{c|}{ACT} & \multicolumn{3}{c|}{ACT $\times$ time} \\[-5pt]
 & & & & & $c$ & $\phi$ & $\eta$	& $c$ & $\gamma$ & $\eta$ \\
\hline
\multirow{4}{*}{10} &     1 &     0.32 &   110000 &     0.20 &      2.5 &      100 &      170 &      0.5 &       20 &       34 \\
	 	 &       10 &     0.32 &    95000 &     0.23 &      2.4 &       91 &      140 &     0.55 &       21 &       32 \\
	 	 &       30 &     0.32 &    80000 &     0.27 &      2.5 &       97 &      150 &     0.68 &       26 &       41 \\
	 	 &       50 &     0.32 &    70000 &     0.30 &      2.7 &       90 &      140 &     0.81 &       27 &       42 \\
\hline 
\multirow{4}{*}{30} &     1 &     0.33 &    80000 &     0.26 &      2.3 &       34 &       68 &      0.6 &      8.8 &       18 \\
	 	 &       10 &     0.33 &    70000 &     0.31 &      2.3 &       18 &       26 &     0.71 &      5.6 &      8.1 \\
	 	 &       30 &     0.33 &    55000 &     0.39 &        2 &       18 &       27 &     0.78 &        7 &       11 \\
	 	 &       50 &     0.34 &    35000 &     0.61 &      2.4 &       12 &       19 &      1.5 &      7.3 &       12 \\
\hline 
\multirow{4}{*}{50} &     1 &     0.34 &    60000 &     0.36 &      1.7 &       33 &       69 &     0.61 &       12 &       25 \\
	 	 &       10 &     0.35 &    50000 &     0.44 &      2.1 &       10 &       15 &     0.92 &      4.4 &      6.6 \\
	 	 &       30 &     0.34 &    30000 &     0.71 &      1.8 &       10 &       15 &      1.3 &      7.1 &       11 \\
	 	 &       50 &     0.34 &    25000 &     0.81 &      1.8 &       12 &       17 &      1.5 &      9.7 &       14 \\
\hline 
\multirow{4}{*}{70} &     1 &     0.34 &    45000 &     0.49 &      2.2 &       29 &       61 &      1.1 &       14 &       30 \\
	 	 &       10 &     0.35 &    30000 &     0.72 &      1.6 &      7.3 &       11 &      1.2 &      5.3 &      7.9 \\
	 	 &       30 &     0.36 &    25000 &     0.86 &      1.6 &      7.3 &      9.3 &      1.4 &      6.3 &        8 \\
	 	 &       50 &     0.36 &    25000 &     0.96 &      1.8 &      5.9 &      7.8 &      1.7 &      5.7 &      7.5 \\
\hline 
\end{tabular}
\caption{Performance of method ENS2.}
\label{table:ens2act}
\end{table}

We ran the KF method for $140,000$ iterations, with estimated autocorrelation times using the original (unweighed) sequence for $(c, \gamma, \eta)$ of $(2.1, 37, 73)$, which after adjusting by computation time of $0.16$ seconds per iteration are $(0.34, 5.9, 12)$. It follows that the ENS1 method with $L_x$ set to $50$ and $L_{\eta}$ set to $10$ is better than the KF method by a factor of about $3.1$ for the parameter $\eta$. For ENS2, the same settings $L_x = 50$ and $L_{\eta} = 10$ appears to give the best results, with ENS2 worse by a factor of about $1.7$ than ENS1 for sampling $\eta$. We also see that the ENS1 and ENS2 methods aren't too sensitive to the particular tuning parameters, so long at there is a sufficient number of ensemble elements both for $x_{i}$ and for $\eta$.

The results show that using a small ensemble ($10$ or so pool states) over $\eta$ is particularly helpful. One reason for this improvement is the ability to use the caching technique to make these updates computationally cheap. A more basic reason is that updates of $\eta$ consider the entire collection of latent sequences, which allows us to make large changes to $\eta$, compared to the Metropolis updates. 

Even though the ENS2 method in this case is outperformed by the ENS1 method, we have only applied it to one data set and there is much room for further tuning and improvement of the methods. A possible explanation for the lack of substantial performance gain with the ensemble method is that conditional on a single sequence, the distribution of $\phi$ has standard deviation comparable to its marginal standard deviation, which means that we can't move too much further with an ensemble update than we do with our Metropolis updates. An indication of this comes from the acceptance rate for ensemble updates of $\gamma$ in ENS2, which we can see isn't improved by much as more pool states are added. 

Parameter estimates for the best performing KF, ENS1 and ENS2 settings are presented in Table \ref{table:est}. These estimates were obtained by averaging samples from all $5$ runs with $10\%$ of the sample discarded as burn-in. We see that the differences between the standard errors are in approximate agreement with the differences in autocorrelation times for the different methods.  

\begin{table}[t]
\small
\centering
\def\arraystretch{1.2}
\begin{tabular}{|c|c|c|c|}
\hline
Method & $c$ & $\gamma$ & $\eta$ \\
\hline
KF & 0.2300   ($\pm$ 0.0004) & 4.3265 ($\pm$ 0.0053) & -3.7015 ($\pm$ 0.0054) \\
ENS1 & 0.2311 ($\pm$ 0.0006) & 4.3228 ($\pm$ 0.0017) & -3.6986 ($\pm$ 0.0015) \\
ENS2 & 0.2306 ($\pm$ 0.0008) & 4.3303 ($\pm$ 0.0025) & -3.7034 ($\pm$ 0.0021) \\
\hline
\end{tabular}
\caption{Estimates of posterior means, with standard errors of posterior means shown in brackets.}
\label{table:est}
\end{table}

\section{Conclusion}

We found that noticeable performance gains can be obtained by using ensemble MCMC based sampling methods for the stochastic volatility model. It may be possible to obtain even larger gains on different data sets, and with even better tuning. In particular, it is possible that the method of updating $\phi$ with an ensemble, or some variation of it, actually performs better than a single sequence method in some other instance. 

The method of Kastner and Fruwirth-Schnatter (2014) relies on the assumption that the state process is linear and Gaussian, which enables efficient state sequence sampling using Kalman filters. The method would not be applicable if this was not the case. However, the ensemble method could still be applied to this case as well. It would be of interest to investigate the performance of ensemble methods for stochastic volatility models with different noise structures for the latent process. It would also be interesting to compare the performance of the ensemble MCMC method with the PMCMC-based methods of Andrieu et. al (2010) and also to see whether techniques used to improve PMCMC methods can be used to improve ensemble methods and vice versa.

Multivariate versions of stochastic volatility models, for example those considered in Scharth and Kohn (2013) are another class of models for which inference is difficult, and that it would be interesting to apply the ensemble MCMC method to. We have done preliminary experiments applying ensemble methods to multivariate stochastic volatility models, with promising results. For these models, even though the latent process is linear and Gaussian, due to a non-constant covariance matrix the observation process does not have a simple and precise mixture of Gaussians approximation. 

\section*{Acknowledgements}

This research was supported by the Natural Sciences and Engineering Research Council of Canada.  A.~S.\ is in part funded by an NSERC Postgraduate Scholarship. R.~N.\ holds a Canada Research Chair in Statistics and Machine Learning.

\section*{References}

\leftmargini 0.2in
\labelsep 0in

\begin{description}

\item
Andrieu, C., Doucet, A. and Holenstein, R. (2010). ``Particle Markov chain Monte Carlo methods'', {\em Journal of the Royal Statistical Society B}, vol.~72, pp.~269-342.

\item 
Kastner, G. and Fruhwirth-Schnatter, S. (2014). ``Ancillarity-sufficiency interweaving strategy (ASIS) for boosting MCMC estimation of stochastic volatility models'', {\em Computational Statistics \& Data Analysis}, vol.~76,  pp.~408-423.

\item
Kim, S., Shephard, N. and Chib, S. (1998). ``Stochastic volatility: likelihood inference and comparison with ARCH models'', {\em Review of Economic Studies}. vol.~65, pp.~361-393.

\item
Lindsten, F. and Schon, T. B.  (2013). ``Backward simulation methods for Monte Carlo statistical inference'', {\em Foundations and Trends in Machine Learning}. vol.~6(1), pp.~1-143.

\item 
Liu, J.S. and Sabatti, C. (2000). ``Generalized Gibbs sampler and multigrid Monte Carlo for Bayesian computation'', {\em Biometrika}, vol.~ 87, pp.~ 353-369. 

\item
Liu, J.S. and Wu, Y.N. (1999). ``Parameter expansion for data augmentation'', {\em Journal of the American Statistical Association} vol.~94, pp.~1264-1274. 

\item
Neal, R. M. (2003). ``Markov Chain Sampling for Non-linear State Space Models using Embedded Hidden Markov Models'', Technical Report No. 0304, Department of Statistics, University of Toronto, http://arxiv.org/abs/math/0305039.

\item
Neal, R. M., Beal, M. J., and Roweis, S. T. (2004). ``Inferring state sequences for non-linear systems with embedded hidden Markov models'', in S. Thrun, et al (editors), {\em Advances in Neural Information Processing Systems 16}, MIT Press.

\item
Neal, R. M. (2010). ``MCMC Using Ensembles of States for Problems with Fast and Slow Variables such as Gaussian Process Regression'', Technical Report No. 1011, Department of Statistics, University of Toronto,
 http://arxiv.org/abs/1101.0387.

\item
Omori, Y., Chib, S., Shephard, N. and Nakajima, J. (2007). ''Stochastic volatility model with leverage: fast and efficient likelihood inference'', {\em Journal of Econometrics}, vol.~140-2, pp.~425-449.

\item
Petris, G., Petrone, S. and Campagnoli, P. (2009). \textit{Dynamic Linear Models with R}, Springer: New York.

\item 
Scharth, M. and Kohn, R. (2013). ``Particle Efficient Importance Sampling'', arXiv preprint 1309.6745v1.

\item
Shestopaloff, A. Y. and Neal, R. M. (2013). ``MCMC for non-linear state space models using ensembles of latent sequences'', Technical Report, http://arxiv.org/abs/1305.0320.

\item
Yu, Y. and Meng, X. (2011). ``To Center or Not to Center, That is Not the Question: An Ancillarity-Sufficiency Interweaving Strategy (ASIS) for Boosting MCMC Efficiency'', {\em Journal of Computational and Graphical Statistics}, vol.~20 (2011), pp.~531-570.

\end{description}

\section*{Appendix}

Here, we derive the sufficient statistics for the stochastic volatility model in the two parametrizations and the likelihoods in terms of sufficient statistics.

For NC, we derive low-dimensional sufficient statistics for $\phi$ as follows
\begin{eqnarray*}
p(x|\phi) &\propto& \sqrt{1 - \phi^{2}}\exp\Big(-(1 - \phi^{2})x_{1}^{2}/2)\exp(-\sum_{i=2}^{N}(x_{i} - \phi x_{i-1})^{2}/2\Big) \\
&\propto& \sqrt{1 - \phi^{2}}\exp\Big(-(x_{1}^{2} - \phi^{2}x_{1}^{2} + \sum_{i=2}^{N}x_{i}^{2} - 2\phi\sum_{i=2}^{N}x_{i}x_{i-1} + \phi^{2}\sum_{i=2}^{N}x_{i-1}^{2})/2\Big) \\
&\propto& \sqrt{1 - \phi^{2}}\exp\Big(-(\phi^{2}\sum_{i=2}^{N-1}x_{i}^{2} - 2\phi\sum_{i=2}^{N}x_{i}x_{i-1} + \sum_{i=1}^{N}x_{i}^{2})/2\Big)
\end{eqnarray*}
Letting
\begin{eqnarray*}
t_{1} = \sum_{i=1}^{N}x_{i}^{2}, \quad t_{2} = \sum_{i=2}^{N}x_{i-1}x_{i}, \quad t_{3} = x_{1}^{2} + x_{N}^{2}
\end{eqnarray*}
we can write 
\begin{eqnarray*}
p(x | \phi) &\propto& \sqrt{1 - \phi^{2}}\exp(-(\phi^{2}(t_{1} - t_{3}) - 2\phi t_{2} + t_{1})/2)
\end{eqnarray*}
For C, we have
\begin{eqnarray*}
p(x | c, \phi, \sigma^{2}) &\propto& \frac{\sqrt{1 - \phi^{2}}}{\sigma^{n/2}}\exp\Big(-(1 - \phi^{2})(\tilde{x}_{1} - c)^2/2\sigma^{2})\exp\Big(-\sum_{i=2}^{N}((\tilde{x}_{i} - (c + \phi(\tilde{x}_{i-1} - c))^{2}/2\sigma^{2}\Big) \\
&\propto& \frac{\sqrt{1 - \phi^{2}}}{\sigma^{n/2}}\exp\Big(-(\tilde{x}_{1}^{2} - \phi^{2}\tilde{x}_{1}^{2} - 2\tilde{x}_{1}c + c^{2} + 2c\phi^{2}\tilde{x}_{1} - c^{2}\phi^{2} + \sum_{i=2}^{N}\tilde{x}_{i}^{2} \\ 
&&\ \ \ \ - 2\sum_{i=2}^{N}\tilde{x}_{i}(c + \phi(\tilde{x}_{i-1} - c)) + \sum_{i=2}^{N}(c + \phi(\tilde{x}_{i-1} - c))^{2})/2\sigma^{2}\Big) \\
&\propto& \frac{\sqrt{1 - \phi^{2}}}{\sigma^{n/2}}\exp\Big(-(\sum_{i=1}^{N}\tilde{x}_{i}^{2} - \phi^{2}\tilde{x}_{1}^{2} - 2\tilde{x}_{1}c + c^{2} + 2c\phi^{2}\tilde{x}_{1} - c^{2}\phi^{2} - 2c\sum_{i=2}^{N}\tilde{x}_{i}  \\ 
&&\ \ \ \ - 2\phi\sum_{i=2}^{N}\tilde{x}_{i}\tilde{x}_{i-1} + 2c\phi\sum_{i=2}^{N}\tilde{x}_{i} + (N-1)c^{2} + 2c\phi\sum_{i=2}^{N}(\tilde{x}_{i-1} - c) + \phi^{2}\sum_{i=2}^{N}(\tilde{x}_{i-1} - c)^{2})/2\sigma^{2}\Big) \\
&\propto& \frac{\sqrt{1 - \phi^{2}}}{\sigma^{n/2}}\exp\Big(-(\sum_{i=1}^{N}\tilde{x}_{i}^{2} - \phi^{2}\tilde{x}_{1}^{2} - 2\tilde{x}_{1}c + c^{2} + 2c\phi^{2}\tilde{x}_{1} - c^{2}\phi^{2} - 2c\sum_{i=2}^{N}\tilde{x}_{i} \\
&&\ \ \ \ - 2\phi\sum_{i=2}^{N}\tilde{x}_{i}\tilde{x}_{i-1} + 2c\phi\sum_{i=2}^{N}\tilde{x}_{i} + (N-1)c^{2} + 2c\phi\sum_{i=2}^{N}\tilde{x}_{i-1} - 2(N-1)c^{2}\phi) \\ 
&&\ \ \ \ + \phi^{2}\sum_{i=2}^{N}\tilde{x}_{i-1}^{2} - 2c\phi^{2}\sum_{i=2}^{N}\tilde{x}_{i-1} + (N-1)c^{2}\phi^{2})/2\sigma^{2}\Big) \\
&\propto&\frac{\sqrt{1 - \phi^{2}}}{\sigma^{n/2}}\exp\Big(-(\sum_{i=1}^{N}\tilde{x}_{i}^{2} + \phi^{2}\sum_{i=2}^{N-1}\tilde{x}_{i}^{2} - 2\phi\sum_{i=2}^{N}\tilde{x}_{i-1}\tilde{x}_{i} + 2c\phi^{2}\sum_{i=2}^{N-1}\tilde{x}_{i} - 2c\sum_{i=1}^{N}\tilde{x}_{i}  \\
&&\ \ \ \ + 4c\phi\sum_{i=2}^{N-1}\tilde{x}_{i} + 2c\phi(\tilde{x}_{1} + \tilde{x}_{N}) + (N-1)(c(\phi - 1))^{2} + c^{2}(1 - \phi^{2}))/2\sigma^{2}\Big)
\end{eqnarray*}
Letting
\begin{eqnarray*}
\tilde{t}_{1} = \sum_{i=1}^{N}\tilde{x}_{i}^{2}, \quad \tilde{t}_{2} = \sum_{i=2}^{N-1}\tilde{x}_{i}^{2} \quad \tilde{t}_{3} = \sum_{i=2}^{N}\tilde{x}_{i-1}\tilde{x}_{i} \quad \tilde{t}_{4} = \sum_{i=2}^{N-1}\tilde{x}_{i} \quad \tilde{t}_{5} = \tilde{x}_{1} + \tilde{x}_{N}
\end{eqnarray*}
we can write
\begin{eqnarray*}
p(x | c, \phi, \sigma^{2}) &=& \frac{\sqrt{1 - \phi^{2}}}{\sigma^{n/2}}\exp\Big(-(\tilde{t}_{1} + \phi^{2}\tilde{t}_{2} - 2\phi \tilde{t}_{3} - 2c\phi^{2} \tilde{t}_{4} - 2c(\tilde{t}_{4} + \tilde{t}_{5}) \\ 
&&\ \ \ \ + 4c\phi \tilde{t}_{4} + 2c\phi\tilde{t}_{5} + (N-1)(c(\phi -1))^{2} + c^{2}(1-\phi^{2}))/2\sigma^{2}\Big)
\end{eqnarray*}
\end{document}